\documentclass{Interspeech2024}
\usepackage{multirow}

\interspeechcameraready 

\title{Bird Vocalization Embedding Extraction Using Self-Supervised \\ Disentangled Representation Learning}

\name[affiliation={1}]{Runwu}{Shi}
\name[affiliation={2}]{Katsutoshi}{Itoyama}
\name[affiliation={1}]{Kazuhiro}{Nakadai}

\address{
  $^1$Tokyo Institute of Technology, Japan\\
  $^2$Honda Research Institute Japan, Co., Ltd., Japan}
\email{\{shirunwu, itoyama, nakadai\}@ra.sc.e.titech.ac.jp}


\begin{document}

\maketitle

\begin{abstract}
 % To extend this, i
 This paper addresses the extraction of the bird vocalization embedding from the whole song level using disentangled representation learning (DRL). Bird vocalization embeddings are necessary for large-scale bioacoustic tasks, and self-supervised methods such as Variational Autoencoder (VAE) have shown their performance in extracting such low-dimensional embeddings from vocalization segments on the note or syllable level. To extend the processing level to the entire song instead of cutting into segments, this paper regards each vocalization as the generalized and discriminative part and uses two encoders to learn these two parts. The proposed method is evaluated on the Great Tits dataset according to the clustering performance, and the results outperform the compared pre-trained models and vanilla VAE. Finally, this paper analyzes the informative part of the embedding, further compresses its dimension, and explains the disentangled performance of bird vocalizations.
\end{abstract}

\section{Introduction}

Automatic bioacoustic analysis requires the collection of various vocalizations within one species. Such vocal repertoires facilitate the diversity analysis of vocalization and quantitative analysis of vocal behavior. Typically, this process can be conducted by human experts, however, when the categories of repertoire come to hundreds and thousands, this work will be both time consuming and subject to bias. Such situations provide opportunities for self-supervised methods, which do not require large amounts of annotated data. 

One type of such method uses pre-trained models to extract embeddings of specific layers \cite{Ghani-GlobalBirdsongEmbeddingsEnable-2023,Sarkar-CanSelfSupervisedNeuralRepresentations-2023}. The advantage of this approach favors the case of a limited dataset, where a priori knowledge from other fields can be utilized. The other type considers using self-supervised learning based on the Autoencoder (AE) structure, after which the encoder output will be regarded as suppressed embedding \cite{Suzuki-ExtractingBirdVocalizationsComplex-2023}. In \cite{Best-DeepAudioEmbeddingsVocalisation-2023}, a convolutional auto-encoder network is used to learn the abstract embedding of vocalization segments in 6 species, including birds and marine mammals, and performance is quantitatively evaluated using clustering results. In \cite{Goffinet-LowdimensionalLearnedFeatureSpaces-2021}, the Variational Autoencoder (VAE) is adopted to learn the vocal embedding of syllables from laboratory mouse and zebra finch, and the results prove that such learned features outperform handpicked features in a variety of downstream tasks.  

\begin{figure}[t]
  \centering
  \includegraphics[width=\linewidth]{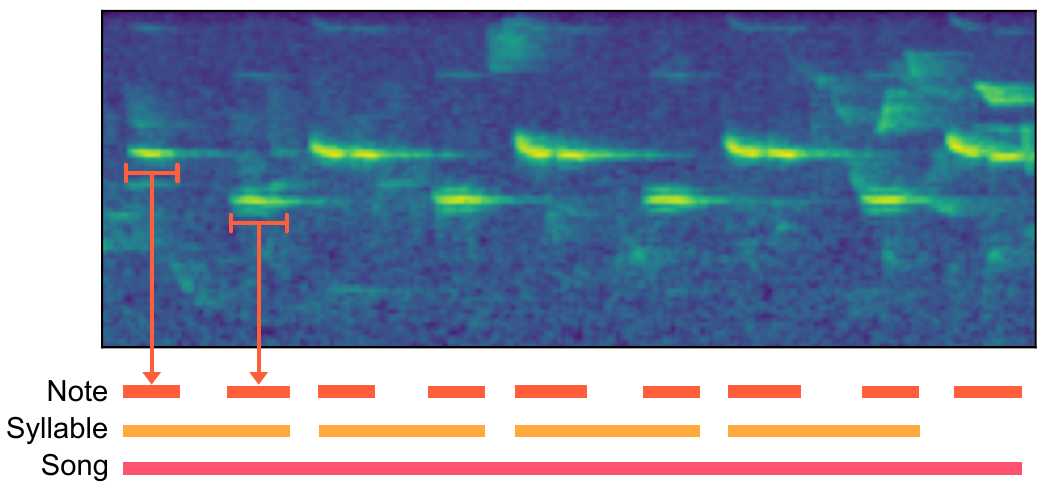}
  \caption{Different elements in bird song (Great Tit).}
  \label{fig:one}
\end{figure}

Despite recent work related to animal vocalization embedding using self-supervised learning has made significant progress, some practical issues remain. Most of the related methods focus on the note or syllable level of vocalization, while some bird songs contain various levels of elements. The songs of Great Tits can be divided into various hierarchical levels as shown in Figure~\ref{fig:one} \cite{Recalde-DenselySampledRichlyAnnotated-2023}, which can be regarded as a special case of sound ontology \cite{Nakatani-SoundOntologyComputationalAuditory-1998}. The note is the most fundamental unit separated by silence. The syllable is the sequence of notes repeated in the same order in a song. Beyond this is the song that consists of the same syllables. Typically, to extract notes or syllables from a continuous song, some methods such as threshold detection are necessary \cite{MerinoRecalde-PykantoPythonLibraryAccelerate-2023}, while it should be noticed that such methods are always sensitive to background noise and the characteristics of different notes, which require researcher's experience and inspection. Moreover, cutting notes inevitably omits original information, and the syntactic relationships among notes are also not fully considered. To build convincing vocalization repertoires on the song level, obtaining the corresponding embeddings directly from the entire song is necessary. 

However, learning song embeddings from the entire song using the vanilla VAE structure is challenging. Firstly, it is common for the same type of song to have different numbers of notes repeated, which leads to different lengths of songs. This information will be mixed with the discriminative information and make the embeddings more ambiguous. For instance, embeddings of songs with different lengths can easily be clustered into different groups. Moreover, songs are prone to have syntax changes at the note level, such as 'A-B-A-B' and 'B-A-B' \cite{Recalde-DenselySampledRichlyAnnotated-2023}, combined with background noise, embeddings should learn to filter out these effects and focus only on critical syntax contents. 

\begin{figure}[t]
  \centering
  \includegraphics[width=\linewidth]{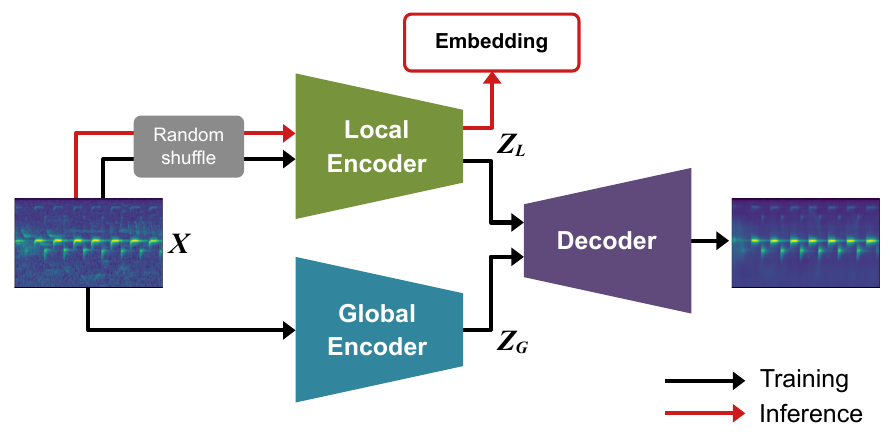}
  \caption{The framework of the proposed method.}
  \label{fig:two}
\end{figure}

Considering such issues, this paper proposes a method to extract song embedding at the song level based on disentangled representation learning (DRL). Instead of using only one encoder, two encoders are adopted to learn the global feature and the local feature simultaneously, in which the global feature represents the temporal related information including fundamental elements such as the number and the position of the notes, and the local feature represents the discriminative information of each song, as shown in Figure~\ref{fig:two}. These local features will be used as vocalization embedding of each song. 
% Ideally, these local features should not contain general fundamental element information, which avoids redundant information and enhances the performance of clustering. 
For evaluation, the proposed method is compared with the embeddings of the pre-trained baseline model on clustering performance, and finally, the interpretation of the learned embeddings is discussed. Our main contributions can be summarized as:

\begin{itemize}
\item The first attempt to disentangle the structured bird vocalization, and the disentangled embeddings presents a better performance.

\item Analyze the information amount learned by embeddings and give the interpretation and embedding compression method.
\end{itemize}

% The rest of the paper is organized as follows. Section 2 explains the dataset and method used in detail. Section 3 gives the experiment steps and the results. Section 4 presents the analysis and discussion. Section 5 gives the conclusion.

\section{Proposed method}

% This section introduce the dataset and the method used to extract vocalization embeddings. The scale of the dataset and preprocessing of each acoustic signal segment are described, after which is the model structure and training strategy. 
This section introduces the proposed method including the modeling motivation, model structure, and training strategy.

% \begin{figure}[t]
%   \centering
%   \includegraphics[width=\linewidth]{figure.pdf}
%   \caption{Length distribution of dataset.}
%   \label{fig:speech_production}
% \end{figure}

\subsection{Structure}

Most of the Great Tits' songs consist of repeated fundamental elements such as notes and syllables, beyond this, this paper considers that the features of such songs can also be divided into global features and local features, and such structured songs should also be the prerequisite for disentanglement. The overall framework with two separate encoders is shown in Figure~\ref{fig:two}.

The global features could be shared among different types of songs and different bird individuals such as the shape of each note, the repetition times related to the length of the song, and even the temporal position of background noise, etc. More importantly, such global features should not contain information related to discrimination, as it's really common for different types of songs to share the same global feature such as the same times of repeated notes. Denote the dataset $X=\{x_{1},x_{2},...,x_{N}\}$ consisting of $N$ Mel-spectrogram of each song segment, and denote $Z_{G}\in R^{d_{G}\times T}$ as the global latent representation, where $T$ is the length of the spectrogram. Given the trainable parameters of encoders $\phi$, the global feature can be written using posterior distribution as $q_{\phi} (Z_{G} \vert X)$.

For local features, such features should contain more discriminative information that can distinguish diverse songs from different individuals. For instance, the spectral information on the note level and the relationship among the nearby notes representing the syntax content should be contained in these features. This kind of high-level representation should be extracted from the original song inputs and separated from the global features that encode more generalized information. Such features with less redundant information should significantly improve the performance on downstream tasks such as clustering embeddings. The local features are one-dimensional vectors, denoted as $Z_{L}\in\mathbb{R}^{d_{L}}$, which can be represented using posterior distribution as $q_{\phi} (Z_{L}\vert X)$.

This idea of modeling comes from disentangled representation learning (DRL) for human speech and musical signals \cite{Lu-SpeechTripleNetEndtoEndDisentangledSpeech-2023, Lian-RobustDisentangledVariationalSpeech-2022a, Polyak-SpeechResynthesisDiscreteDisentangled-2021, Tanaka-PitchTimbreDisentanglementMusicalInstrument-2021, Wu-SelfSupervisedDisentanglementHarmonicRhythmic-2023}. The DRL can be defined as the learning paradigm that aims to obtain representations capable of identifying and disentangling the underlying factors hidden in the observed data \cite{Wang-DisentangledRepresentationLearning-2023}. For instance, the speech can be disentangled into content, speaker, and prosody information. Such methods always assume that in continuous speech, the content information is temporal dependent and the speaker information is temporal invariant \cite{Liu-DisentanglingVoiceContentSelfSupervision-2023, Hsu-UnsupervisedLearningDisentangledInterpretable-2017, Qian-UnsupervisedSpeechDecompositionTriple-2020}, these two kinds of latent information will be learned using two separate encoders under the VAE structure. Similar to these related works, we use a temporal encoder for the global feature, and a down sampling encoder with randomly shuffled spectrogram input for the local feature.

The structure of the model is adopted from the SpeechTripleNet \cite{Lu-SpeechTripleNetEndtoEndDisentangledSpeech-2023}. The global encoder uses a 1D convolution layer with kernel size 1 to project the dimension of the input spectrogram into 256, followed by two 1D convolution layers and batch normalization layers. Then two self-attention layers are used to learn the temporal relationship. Finally, global features of shape [128, T] are sampled from a multidimensional Gaussian distribution using the reparameterization trick, where T is the length of the spectrogram. For the local encoder, the input spectrogram will be randomly shuffled with the fixed length segment 32, which can be regarded as ignoring the long term temporal relationship and focusing only on local information \cite{Qian-UnsupervisedSpeechDecompositionTriple-2020}. Then, a 1D convolutional layer with kernel size 1 is used to enlarge the dimension of the spectrogram into 256, followed by three 1D convolutional layers with the kernel size of 3, 3, and 5 respectively, and each convolutional layer is followed by an average pooling layer with the size of 2. After layer normalization, the reparameterization trick is used to sample local features, a one-dimensional vector containing 128 latent units. For the decoding process, the local features are firstly expanded to the same shape as the global feature and concatenated with it and then the concatenated feature with the shape of [256, $T$] is input into the decoder. The decoder has the same structure as the global encoder embedded with self-attention layers that can extract temporal information for better reconstruction. The trainable parameters of the decoder are represented as $\theta$, and the decoder models the conditional probability $p_{\theta}(X \vert Z_{G},Z_{L})$ given the two latent features. The shape of the output is the same as the input spectrogram. 

\subsection{Training strategy}

The training strategy determines the performance of disentangling the useful local features from songs. To achieve better disentanglement, the hyperparameters $\gamma_{G}$, $\gamma_{L}$, $C_{G}$, and $C_{L}$ are adopted in the loss function compared to the original VAE, as shown in equation~\ref{equation:first}. The learning objective is to minimize this, equivalent to maximizing the lower bound of $\log p_{\theta} (X\vert Z)$.

\begin{align}
  \mathcal{L} = & \mathbb{E}_{X,q_{\phi}(Z_G,Z_L \vert X)} [ \log p_{\theta}(X \vert Z_G,Z_L) ] + \nonumber \\ 
  & \mathbb{E}_X [ \gamma_G \vert D_{KL}(q_{\phi}(Z_G \vert X) \parallel p(Z_G)) - C_G \vert] + \nonumber \\
  & \mathbb{E}_X [ \gamma_L \vert D_{KL}(q_{\phi}(Z_L \vert X) \parallel p(Z_L)) - C_L \vert ] 
  \label{equation:first}
\end{align}

where the first term is the reconstruction loss, and the second and the third terms are the KL divergence of global and local latent features. The $\gamma_{G}$ and $\gamma_{L}$ are weight factors that control the disentangled extent, a larger value (larger than 1) always results in a more disentangled latent representation \cite{Burgess-UnderstandingDisentanglingBetaVAE-2018}. In addition, the value of $\gamma_{G}/\gamma_{L}$ also influences the information flow between global and local features \cite{Lu-SpeechTripleNetEndtoEndDisentangledSpeech-2023, Lian-RobustDisentangledVariationalSpeech-2022a}. As shown in equation~\ref{equation:second}, the mutual information among the input and the latent representation $I(X;Z)$ is the lower bound of the KL divergence term, which means that the KL divergence is larger than the information that $Z$ can transmit about the input $X$ \cite{Dupont-LearningDisentangledJointContinuous-2018}. 

\begin{align}
  & \mathbb{E}_X [ D_{KL} (q_{\phi} (Z \mid X) \parallel p(Z)) ] \nonumber \\
  = & \mathbb{E}_{q(Z,X)} \left[ \log \frac{q(Z,X)}{q(Z)p(X)} \right] + \mathbb{E}_{q(Z,X)} \left[ \log \frac{q(Z)}{p(Z)} \right] \nonumber \\
  = & I(X;Z) + D_{KL} \left( q(Z) \parallel p(Z) \right) 
  \label{equation:second}
\end{align}

For instance, if the weight of the global encoder $\gamma_{G}$ is given a much larger value than $\gamma_{L}$, the gradient will be dominated by the global encoder output, resulting in the faster vanishing of $D_{KL} \left( q_{\phi} (Z_G \vert X) \parallel p(Z_G) \right)$, which means there will be less information about input data $X$ learned by $Z_G$, and more information will be encoded in $Z_L$. Typically, the value of $\gamma_{G}/\gamma_{L}$  should be larger than 1 to let the $Z_L$ be more informative. To more explicitly control the encoding capacities of the two encoders, the controllable parameters $C_G$ and $C_L$ are adopted in the learning objective. These two bounds pressure the KL divergence to converge to a certain information capacity \cite{Lu-SpeechTripleNetEndtoEndDisentangledSpeech-2023, Burgess-UnderstandingDisentanglingBetaVAE-2018}. Since the local encoder should learn more discriminative information, its capacity bound $C_L$ is much larger than $C_G$.

Empirically, $\gamma_{G}$ and $\gamma_{L}$ are set to 100 and 10, and $C_G$ and $C_L$ are set to 0.4 and 100, respectively, and we find that the capacity value significantly influences the final performance. During training, the capacity is linearly increased to the maximum value in the first 20K steps \cite{Burgess-UnderstandingDisentanglingBetaVAE-2018, Lu-SpeechTripleNetEndtoEndDisentangledSpeech-2023}. 
% We also observe that only several of the units in the local feature denote the main part of information capacity, which will be discussed in the subsequent section. 
The reconstruction term in equation \ref{equation:first} is realized by the negative log-likelihood among the reconstructed and ground truth samples. The Adam optimizer is used with a learning rate of $1e^{-4}$. The model is trained using a batch size of 64 for a total of 200K training steps.
% All the experiments are conducted on a single Nvidia A100.

\section{Experiment}

This section provides the details of dataset and preprocessing, experiments for model training, visualization of the embeddings, and quantitative performance evaluation and comparison.

\subsection{Dataset}

This paper focuses on the vocalization of the Great Tits (Parus major). Great Tits can produce varied songs that always consist of repeated notes and syllables. 
% We consider such structured songs to be a prerequisite for being disentangled into global features and local features. 
For plenty of diverse samples, this paper uses a publicly available dataset that contains the songs of many different individual Great Tits \cite{Recalde-DenselySampledRichlyAnnotated-2023}. The song of Great Tits has individual specific repertoires, which means that each type of song from different bird individuals should be distinguished. The median amount of songs of each Great Tit individual is 4 and the largest amount is 13. The annotation information contains the individual label and the song label of each bird individual, and the labels are obtained using a semi-supervised method which means the labels are not totally accurate \cite{MerinoRecalde-PykantoPythonLibraryAccelerate-2023}, which is ignored in this paper.

For the preprocessing of each song segment, each song is transformed into a Mel-spectrogram. A Fast Fourier Transformation with a window size of 1024 is adopted, and the window shift is set to 256. The number of Mel filters is set to 80. The sampling rate is \SI{22050}{\hertz}, and the cutoff frequency is set to \SI{1500}{\hertz} and \SI{10000}{\hertz}. To facilitate the training, the length of the spectrogram is limited from 100 to 400. After removing the samples that are too long and too short, there are a total of 98207 song segments used for subsequent experiments.

\subsection{Experiment and Results}
To comprehensively evaluate the performance of the proposed method, we randomly extracted 70\% of the bird individuals as the training dataset, 10\% as the validation dataset, and the remaining 20\% for testing. In detail, the training, validation, and test dataset contains 244 individuals with 1112 song types, 35 individuals with 151 song types, and 71 individuals with 305 song types, respectively, and there are a total of 67425, 10764, and 20018 song segments in these three subdatasets. We evaluate the clustering performance of the embeddings from the individuals that the model has never seen, to verify if the model has truly learned the discriminative representation.

The process of clustering includes embedding extraction, dimension reduction, and clustering. Firstly, all the local features of each song segment will be collected as the song embedding with a length of 128, then the UMAP is used to reduce the dimension of the embeddings \cite{Ghani-GlobalBirdsongEmbeddingsEnable-2023, Goffinet-LowdimensionalLearnedFeatureSpaces-2021, Sainburg-FindingVisualizingQuantifyingLatent-2020b}, and after that, the HDBSCAN algorithm is used to cluster these condensed embeddings, which is always used in bioacoustic field \cite{Best-DeepAudioEmbeddingsVocalisation-2023, Sainburg-FindingVisualizingQuantifyingLatent-2020b}. The label information is only used to check the correctness of the clustering results. The method in \cite{Best-DeepAudioEmbeddingsVocalisation-2023} is used to search for optimized parameters of UMAP and HDBSCAN, resulting in the UMAP unit of 4, cluster size of 5, minimum samples of 3, and epsilon of 0.1.

To quantitatively evaluate the clustering of the obtained embeddings, Normalised Mutual Information (NMI) is used to represent the extent to which embeddings are correctly clustered. Given the clusters $C$ and labels $L$, the NMI calculates the relative entropy between the joint distribution $P_{L,C}$ and the product of $P_L$ and $P_C$, and normalized by the sum of the entropy of $L$ and $C$, as shown in equation~\ref{equation:third}. The NMI will be 1 if the labels match the clusters perfectly. For all compared methods, we use the most optimized clustering parameters to calculate the NMI.

\begin{align}
  NMI(L;C) = \frac{D_{KL} \left( P_{L,C} \parallel P_L \otimes P_C \right) \times 2}{H(L) + H(C)}
  \label{equation:third}
\end{align}

\begin{figure}[t]
  \centering
  \includegraphics[width=\linewidth]{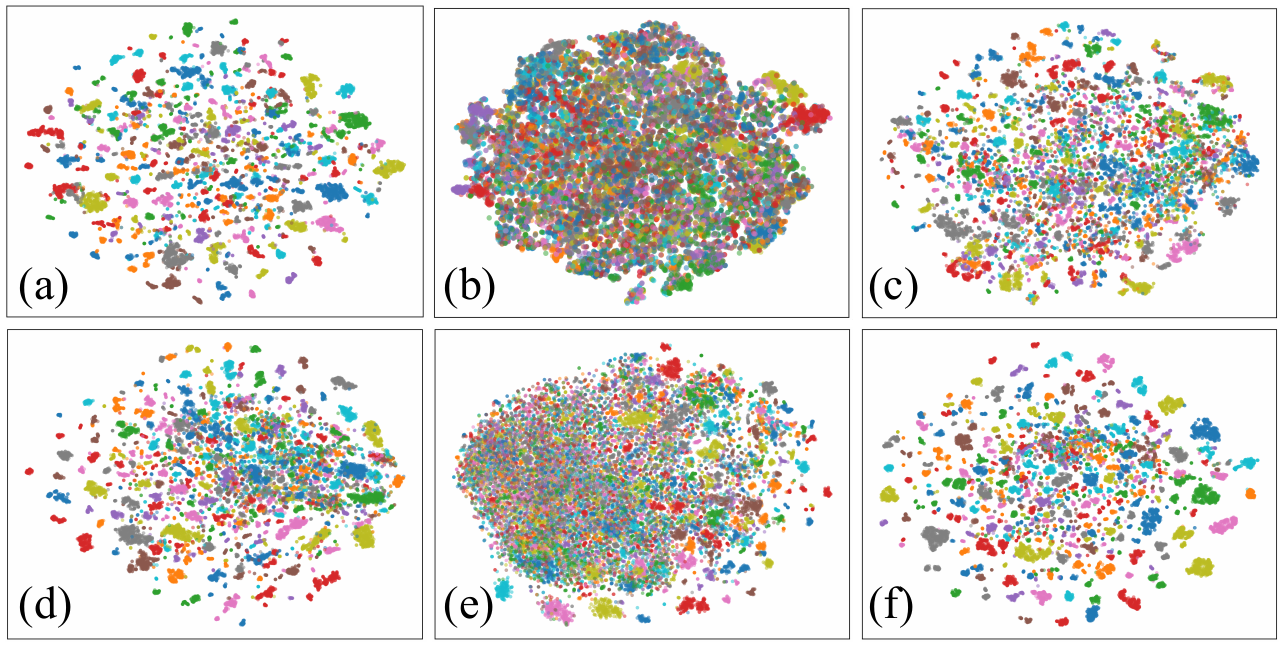}
  \caption{T-SNE of embeddings of compared methods with different colors for different song types. (a) Local encoder of the proposed method, (b) VAE (Global Encoder with Decoder), (c) Wav2Vec2, (d) Hubert, (e) VQ-APC, (f) OpenL3. }
  \label{fig:three}
\end{figure}

% two
% \begin{figure*}[t]
%   \centering
%   \includegraphics[width=\linewidth,height=5cm]{figure.pdf}
%   \caption{T-SNE of embeddings of different methods.}
%   \label{fig:three}
% \end{figure*}

For comparison, we use the embeddings of pre-trained baseline models and the latent representation of vanilla VAE. We choose several baseline models including Wav2Vec2 \cite{Baevski-Wav2vecFrameworkSelfSupervisedLearning-2020}, Hubert \cite{Hsu-HuBERTSelfSupervisedSpeechRepresentation-2021}, VQ-APC \cite{Chung-VectorQuantizedAutoregressivePredictiveCoding-2020}, and OpenL3 \cite{Cramer-LookListenLearnMore-2019}. The first 3 baseline models are trained in human speech and have demonstrated their ability in bioacoustic tasks \cite{Sarkar-CanSelfSupervisedNeuralRepresentations-2023, Best-DeepAudioEmbeddingsVocalisation-2023}. The OpenL3 provides audio embeddings which can be used for acoustic scene classification. For the vanilla VAE \cite{Kingma-AutoEncodingVariationalBayes-2022}, the same structure and the training method as the global encoder is adopted, the only difference being that the embedding channel is compressed to 1 along the time dimension. The T-SNE map of the embeddings is shown in Figure~\ref{fig:three}. Different colors mean different song types. The embedding output of the local encoder clusters different song types more clearly than other methods, proving the local encoder learns more discriminative features. For comparison, the clustering of vanilla VAE using only one encoder is much more ambiguous, and the OpenL3 embeddings perform better than other baselines.

\begin{table}[ht]
  \caption{Performance comparison of different methods.}
  \label{tab:one}
  \centering
  \begin{tabular}{p{1.8cm}p{2cm}p{1.5cm}p{0.9cm}}
    \toprule
    \textbf{Methods} & \textbf{Parameter} & \textbf{Dimension} & \textbf{NMI} $\uparrow$  \\
    \midrule
    \multirow{3}{1cm}{\parbox{2cm}{Ours \\[-2mm] \rule{\linewidth}{0.1pt} \\ \textbf{Ours}(compress)}} 
    & \multirow{3}{1cm}{1.7M(Local~Enc*)\\7.2M(Global~Enc)\\7.3M(Decoder)} 
    & \multirow{3}{1cm}{\parbox{2.9cm}{128 \\[-2mm] \rule{\linewidth}{0.1pt} \\ \textbf{27}}} 
    & \multirow{3}{1cm}{\parbox{1.5cm}{0.901 \\[-2mm]  \\ \textbf{0.902}}} \\
    % & \multirow{3}{*}{\parbox{1cm}{\raisebox{-2.8mm}[0pt][0pt]{\makebox[1cm][r]{0.901}} 
    % & \multirow{3}{*}{\parbox{1cm}{\raisebox{-2.8mm}[0pt][0pt]{\makebox[1cm][r]{0.901}} \vspace{1mm}\\ \raisebox{0.5mm}[0pt][0pt]{\makebox[1cm][r]{0.902}}}}  \\
    & & &  \\
    &&&\\
    \hline
    VAE & 14.5M & 128 & 0.426 \\
    % \cline{1-1} \cline{4-4} % line
    Wav2vec2~\cite{Baevski-Wav2vecFrameworkSelfSupervisedLearning-2020} & 95.0M & 768 & 0.741 \\
    Hubert~\cite{Hsu-HuBERTSelfSupervisedSpeechRepresentation-2021} & 94.7M & 768 & 0.827 \\
    VQ-APC~\cite{Chung-VectorQuantizedAutoregressivePredictiveCoding-2020} & 4.6M & 512 & 0.494 \\    
    OpenL3~\cite{Cramer-LookListenLearnMore-2019} & 4.7M & 6144 & 0.895 \\
    \bottomrule
  \end{tabular}
\end{table}

The NMI results of these methods are shown in Table~\ref{tab:one}. The NMI of the proposed method is 0.901, and after compression, our method obtains a higher score of 0.902 with a much lower dimension, which will be discussed in the next section. For comparison, the embedding of vanilla VAE only has an NMI of 0.426. The Wav2vec2 gets an NMI of 0.741 and the Hubert gets a higher value of 0.827, suggesting that the structural features learned from human speech can also be extended to bird songs. For the OpenL3 trained on video which provides multi-modality information, the NMI comes to an impressive value of 0.895. The OpenL3 trains on a large amount of data including almost 40M samples \cite{Cramer-LookListenLearnMore-2019}. Such diversity of data inputs makes the method highly generalizable to bioacoustic tasks with advanced performance \cite{Best-DeepAudioEmbeddingsVocalisation-2023, Turian-HEARHolisticEvaluationAudio-2022}. 

\section{Analysis and Discussion}

This section analyzes the obtained embeddings and observes the imbalance of the amount of information in different embedding units, from which more informative embedding units can be extracted to dramatically decrease the dimension of embedding.
% , and by adjusting these key units, specific elements in the song spectrogram can be edited, proving the unit truly learn the representation of song features. 

\begin{figure}[t]
  \centering
  \includegraphics[width=\linewidth]{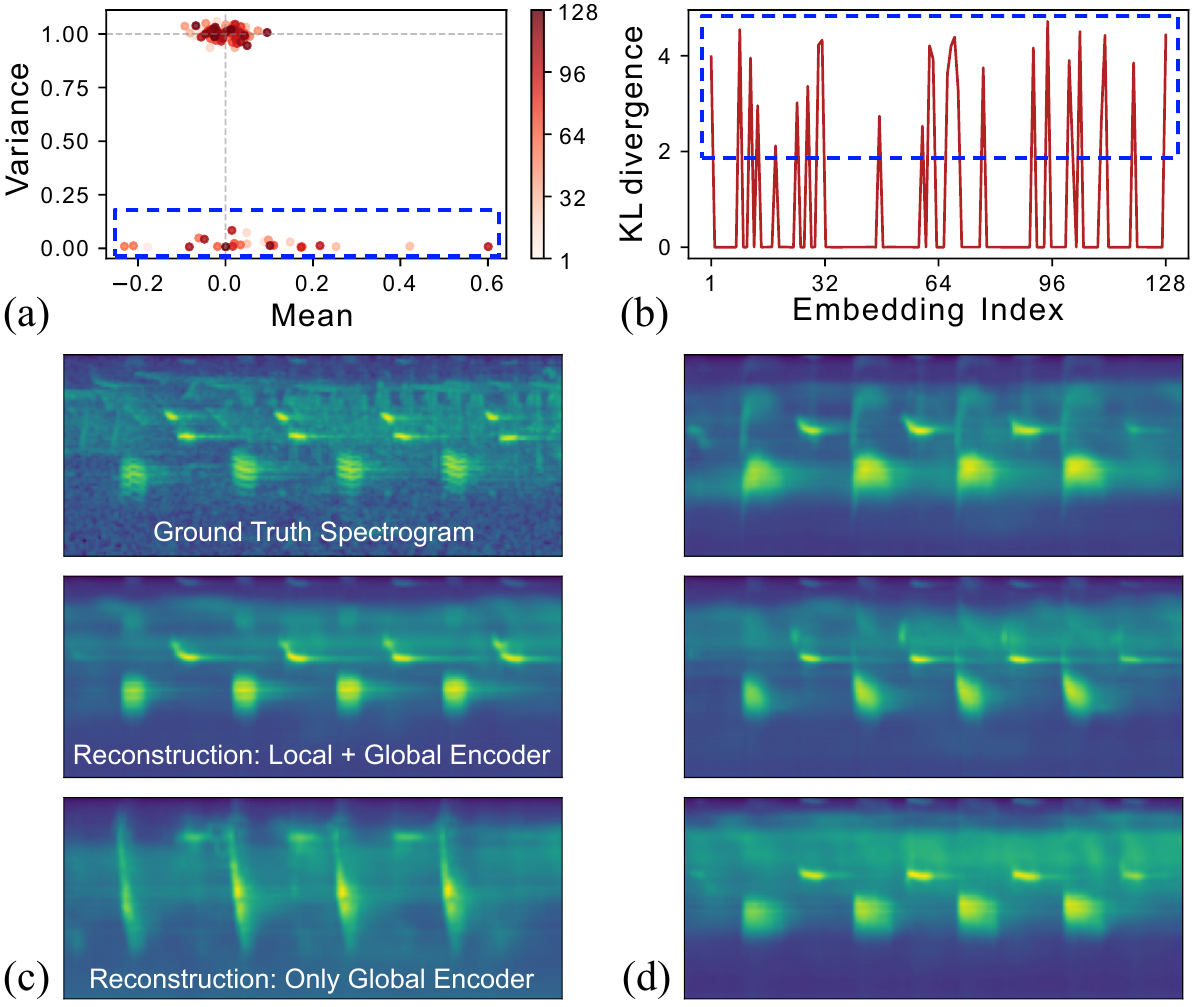}
  \caption{(a) Each unit's element-wise mean and variance, (b) KL divergence of each unit in embedding, (c) Reconstruction results using all or only the global encoder, (d) Reconstruction results when changing informative units in embedding.}
  \label{fig:four}
\end{figure}

Each song embedding with the shape of [1, 128] is sampled from the multidimensional Gaussian distribution with 128 latent units, from which the KL divergence between each unit and the normal distribution can be calculated to estimate the amount of information learned by different units. As shown in Figure~\ref{fig:four} (a), the $x$ and $y$ axis represent the mean and variance of each unit, and the color bar means different unit indexes. In Figure~\ref{fig:four} (b), the $x$ axis represents the unit index and $y$ axis represents the value of KL divergence, and those units with much larger KL divergence correspond to the unit with lower variance and diverse mean values, as shown in the blue box in these two figures. These units can also be regarded as deterministic units since randomness is eliminated and is mainly controlled by the mean value. It should be noted that the KL divergence of these units at these specific positions always has larger values with different samples of input, indicating these units are more informative \cite{Dupont-LearningDisentangledJointContinuous-2018}. To verify this, we extract these 27 informative units and conduct the same test as above. These much shorter embeddings can achieve a higher NMI of 0.902, proving these key units represent discriminative parts of the embedding. Moreover, since the sum of the KL divergence is constrained by the total channel capacity in the learning objective, and empirically, a larger capacity usually leads to more informative units. This method is beneficial for more bioacoustic tasks since the informativeness of the repertoire is diverse in different species. 

% \begin{figure}[t]
%   \centering
%   \includegraphics[width=\linewidth]{figure.pdf}
%   \caption{Reconstruction results when changing informative units.}
%   \label{fig:five}
% \end{figure}

We also conduct reconstruction experiments to verify the knowledge learned by these informative units. In detail, the output of the global encoder remains unchanged, only one informative unit of the overall 128 units is chosen and adjusted, and then the concatenated features are fed into the decoder to reconstruct the spectrogram. Figure~\ref{fig:four} (c) shows the original, reconstructed spectrogram, and reconstructed spectrogram with all 128 units set to 0. This shows that the global encoder learns the temporal information, such as the number and position of each note, but lacks the detailed shape of each note. Each row in figure~\ref{fig:four} (d) presents the reconstruction results when one informative unit is adjusted, in which the discriminative features such as note and overtone are changed. However, due to the inconsistency of the element in spectrograms, the extracted informative units do not have isolated disentangled features such as color, direction, and shape as the vision toy dataset \cite{Higgins-BetaVAELearningBasicVisual-2016}. 
% Nevertheless, the results show that these units can be used as embeddings for each song, and discriminative local features in structured songs of Great Tits can be extracted directly from the whole song input.

\section{Conclusions}
This paper demonstrates the feasibility of using DRL to extract bird vocalization embeddings. By adopting DRL, which utilizes two encoders to capture both global and local features of the songs of Great Tits, we achieve the extraction of embeddings from the whole song level, and the clustering performance surpasses the other methods. Furthermore, the analysis reveals the informativeness contained in embedding units, from which the compressed embeddings can be obtained. This approach enhances our understanding of bird vocalization patterns and provides a potential way to measure the information richness of vocal repertoires in more species in further research.  

\section{Acknowledgements}
This work was supported by JSPS KAKENHI Grant No. JP19KK0260, JP20H00475 and JP23K11160.

% Acknowledgement should only be included in the camera-ready version, not in the version submitted for review.
% The 5th page is reserved exclusively for \red{acknowledgements} and  references. No other content must appear on the 5th page. Appendices, if any, must be within the first 4 pages. The acknowledgments and references may start on an earlier page, if there is space.

% \ifinterspeechfinal
%      The Interspeech 2024 organisers
% \else
%      The authors
% \fi
% would like to thank ISCA and the organising committees of past Interspeech conferences for their help and for kindly providing the previous version of this template.

% \newpage

\bibliographystyle{IEEEtran}
\bibliography{template}

\end{document}